\documentclass[12pt]{article}
\usepackage{graphicx}
\usepackage{dcolumn}% Align table columns on decimal point
\usepackage{bm}% bold math
\usepackage{authblk}% bold math
\usepackage{amsmath,amsthm,amssymb}
\usepackage{fullpage}

\def\Lexp{L_{\exp}}
\def\alg{algorithm}

\def\mexp{matrix exponential}

\def\Fd{Fr\'{e}chet derivative}

\def\lhs{left-hand side}

\def\l{\lambda}

\newcommand{\norm}[1]{\|#1\|}

\newcommand{\R}{\mathbb{R}}
\newcommand{\Rnbyn}{\mathbb{R}^{n \times n}}

\newcommand{\nbyn}{{n \times n}}

\renewcommand{\vec}{\mathop{\mathrm{vec}}}

\newcommand{\Sec}[1]{Section~\ref{sec:#1}}
\newcommand{\Fig}[1]{Figure~\ref{fig:#1}}
\newcommand{\Table}[1]{Table~\ref{tab:#1}}

\newcommand{\Eq}[1]{Eq.\ (\ref{eq:#1})}

\usepackage{latexsym}

\begin{document}

\title{Ranking the importance of nuclear reactions for activation and
transmutation events}
%\author{Wayne Arter$^{a,*}$, J. Guy Morgan${^b}$, Samuel D. Relton${^c}$
%and Nicholas J. Higham${^c}$}
%\address{$^{a}$CCFE, Culham Science Centre, Abingdon, Oxon. OX14 3DB, UK. wayne.arter@ukaea.uk\\
%${^b}$Culham Electromagnetics Ltd, Culham Science Centre, Abingdon, Oxon. OX14 3DB\\
%${^c}$School of Mathematics, The University of Manchester, Manchester
%M13 9PL}
\author{Wayne Arter}
\affil{CCFE, Culham Science Centre, Abingdon, Oxon. OX14 3DB, UK. wayne.arter@ukaea.uk}
\author{J. Guy Morgan}
\affil{Culham Electromagnetics Ltd, Culham Science Centre, Abingdon, Oxon. OX14 3DB}
\author{Samuel D. Relton}
\affil{School of Mathematics, The University of Manchester, Manchester 
M13 9PL}
\author{Nicholas J. Higham}
\affil{School of Mathematics, The University of Manchester, Manchester
M13 9PL}

\date{\today}

\begin{abstract}
Pathways-reduced analysis is one of the techniques used by the \textsc{Fispact-II}
nuclear activation and transmutation software to study the sensitivity of the computed
inventories to uncertainty in reaction cross-sections. Although
deciding which pathways are most important is very helpful in for example
determining which nuclear data would benefit from further refinement,
pathways-reduced analysis need not necessarily define the most critical
reaction, since one reaction may contribute to several different pathways.
This work examines three different techniques for ranking reactions
in their order of importance for determining the final inventory,
comparing the pathways-based metric~(PBM), the direct method and one based
on the Pearson correlation coefficient. Reasons why the PBM is to be
preferred are presented.
\end{abstract}

%\pacs{52.35.-g, 47.55.P-, 02.30.Oz}

\maketitle

\section{Introduction}\label{sec:intro}
\textsc{Fispact-II} is a software suite for the analysis of nuclear
activation and transmutation events of all kinds~\cite{ccfe-r11-11}.
The present work focuses on its use in sensitivity studies
of nuclear inventory calculations; these employ the Bateman model for the evolution of the
inventory of a target subject to irradiation  by an imposed
flux of projectile particles, always neutrons herein.
In ref~\cite{Ar14Sens} %Arter and Morgan(2014)
it was established that the pathways-reduced approach~\cite{EaMo13,Ea15Inve} %Eastwood et al (2015)
to such studies, almost invariably gives very close agreement with Monte-Carlo sensitivities
computed using full Bateman, i.e. accounting for all nuclides and pathways.
Pathways-reduced models are, following Eastwood and Morgan~\cite{EaMo13},
identified by a graph-based approach which determines
the key reaction pathways determining the inventory at a given time
and eliminates from consideration those nuclides which do not lie on
this reduced set of pathways.

The pathways-reduced metric is a sensitivity method in the
respect that implicitly it selects a set of the most important nuclide reactions. A wide
range of other sensitivity methods has been used by the nuclear industry
as shown by literature reviews by Helton~et~al~\cite{He06Surv}, see also
Cacuci and Ionescu-Bujor~\cite{Io04Comp,Ca04Comp}.
Indeed, sensitivity analyses are available as part of nuclear industry software packages such as
for example DAKOTA~\cite{dakota6} and SCALE~\cite{scale6}.
General purpose sensitivity software is also available, such as OpenCossan~\cite{Pa14Open}.

A key input to most techniques considered herein is an estimate of the uncertainty in the reaction
cross-section. The determination of such uncertainties is a challenging
subject in its own right, hence it is important to examine techniques that can identify
which reactions most require further examination.
The present work represents a comparison of three different techniques that exploit the
pathways-based reduction for the nuclear activation problem.

\textsc{Fispact-II} can access uncertainty data, typically the standard deviation,
for the vast majority of reaction cross-sections in the EASY-II database~\cite{easy},
however no information is currently passed
concerning pure decay reactions. This reflects the fact that half-lives are often
very accurately known. There are other reactions in the database for which a value
of zero uncertainty is found, usually indicating that no information is available.
There are thus difficulties in making the comparison, the implications of which
are discussed in \Sec{comment}.

To proceed further with this introduction, it is efficient to introduce the time evolution
(rate or Bateman equation) for a nuclear inventory~$X$ 
\begin{equation}\label{eq:rateq}
\frac{dX}{dt}=\mathsf{A}X
\end{equation}
where $X$ is the vector of nuclide numbers, and
$\mathsf{A}$ is the matrix of nuclear interaction coefficients for
both induced reactions and spontaneous decays. Note that
one coefficient $A_{ij}$ of~$\sf{A}$ may represent several different nuclear 
reactions, since the equation involves an average over a spectrum of
energies (of  neutrons in the present work, although other elementary particles
may be considered in general). Hence the term `interaction' is used to
cover all effects generating nuclide~$i$ as the child of parent~$j$.
It is worth noting that although $i$ precedes~$j$ alphabetically,
reactions throughout this work will except for the~$A_{ij}$ 
be described in parent-child order.
In general, the coefficients~$A_{ij}$ may change with time as
the incident neutron flux changes.

All the techniques for ranking the interactions~$A_{ij}$ are most
easily understood in the context of a single constant irradiation
in the time interval~$(0,t_f)$, producing an inventory~$X(t_f)$.
Different aspects of the inventory, such as heat production or kerma,
may be studied using \textsc{Fispact-II}, but for illustrative purposes
it is sufficient to consider only the total activity
\begin{equation}\label{eq:Q}
Q=\sum_{k}\lambda_k X_k(t_f)
\end{equation}
where~$\lambda_k$ is the decay rate of the nuclide~$X_k$; $\lambda_k$
is zero for stable nuclides and $\lambda_k = \log_e2/\tau_k$ for
unstable ones, where $\tau_k$ is the half-life. 
Although attention focuses here on \textsc{Fispact-II}, the Bateman
approach is standard in that most packages with a claim
to generality, not only DAKOTA and SCALE in the U.S. but also ANSWERS~\cite{answers}
with FISPIN in the U.K., include solvers for the problem. The
pathways-based analysis technique studied here could in principle be implemented
in any of these codes.

The three different ranking techniques are described in the
next~\Sec{measures}. Apart from the use of pathways-based reduction,
there is novelty in the calculation of the direct sensitivity, in that the
matrix Fr\'{e}chet derivative is used in its computation, see Appendix,
rather than the more usual decoupled direct method DDM of Dunker~\cite{Du81Effi}.
The application of the techniques to the wide range of test cases 
first introduced in ref~\cite{Ar14Sens} %Arter and Morgan(2014)
is illustrated in \Sec{calculations}. Lastly \Sec{conc} compares
the utility of the different techniques.

\section{Sensitivity Measures}\label{sec:measures}
\subsection{Pathways Based Metric}\label{sec:PBM}
The Pathways Based Metric (PBM) is calculated quite simply from
the output of the pathways-reduced approach, which includes a listing
of each pathway and its percentage contribution to the active nuclide
at its termination. For a given interaction~$A_{ij}$, all the number~$N_p$
of pathways upon which it lies are identified and the PBM calculated as
\begin{equation} \label{eq:PBM}
S_{PBM}^{ij} = \sum_{k=1}^{N_p}p_l \lambda_t X_t I_{kl}
\end{equation}
where~$p_l$ is the fractional contribution of pathway~$l$ to the number
of atoms~$X_t$ (evaluated at time~$t_f$) in the inventory
with decay rate~$\lambda_t$ and
the indicator matrix~$I_{kl}=1$~or~$0$ depending whether or not
a reaction contributing to the interaction lies on the pathway.

\begin{figure}[htbp]
\centerline{\rotatebox{0}{\includegraphics[width=9.0cm]{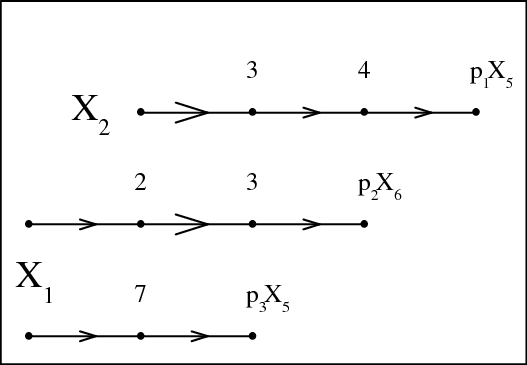}}}
\caption{\label{fig:npath}
Example illustrating how 
the Pathways Based Metric~$S_{PBM}$ is calculated when the three pathways shown
are highlighted by the pathways-reduced analysis. The pathways are numbered in order
$1$, $2$ and~$3$ from the top. Pathway~$1$ starts with nuclide~$2$ and generates
via a sequence of reactions involving nuclides~$3$ and~$4$, $p_1X_5$
atoms of nuclide~$5$, whereas the other pathways begin with nuclide~$1$
and generate nuclides~$6$ and $5$ respectively.
The larger arrowheads indicate reactions with the coefficient of interest $A_{32}$.}
\end{figure}

\Fig{npath} illustrates how the definition works in a simplified case
where irradiation of an initial sample consisting of $X_1$ atoms of nuclide~$1$
and $X_2$ atoms of nuclide~$2$
produces an inventory containing numbers $X_5$ and~$X_6$ of
radioactive nuclides $5$ and~$6$ respectively, with $3$~important pathways.
The first pathway contributes $p_1X_5$ atoms and the third $p_3X_5$
atoms of nuclide~$5$.
(Supposing that other pathways are unimportant, $p_1+p_3\approx 1$ and $p_2 \approx 1$.)
The sensitivity of the inventory to the reaction with coefficient $A_{32}$
(large arrowheads in \Fig{npath}), is then
\begin{equation} \label{eq:PBMeg}
S_{PBM}^{32} = p_1 \lambda_5 X_5 + p_2 \lambda_6 X_6 
\end{equation}
where $\lambda_5$ is the decay rate of nuclide~$5$ etc.
%\clearpage

This technique required special modification to \textsc{Fispact-II} for its
implementation, which was facilitated by the object-oriented design
of the Fortran-95 code.
For the purposes of initial investigation, the loops which are
identified by the graph-based approach used by \textsc{Fispact-II}
are ignored.

\subsection{Direct Method}\label{sec:DM}
The Direct Method (DM) %%after Dunker(1981)~\cite{Du81Effi}
works directly with the tensor
describing the rate of variation of the nuclide~$X_k$ with respect
to nuclear reaction coefficients. For initial investigative purposes it is
sufficient to consider the partial derivative with respect to~$A_{ij}$.
Differentiating \Eq{rateq} with $(i,j)$ regarded as fixed, gives
\begin{equation} \label{eq:ddmeq}
\frac{d}{dt}\left(\frac{\partial X}{\partial A_{ij}}\right) =
A \frac{\partial X}{\partial A_{ij}} + \frac{\partial A}{\partial A_{ij}} X
\end{equation}
If the sensitivity of the total activity is required, then using \Eq{Q}, this is
\begin{equation} \label{eq:SDM}
S_{DM}^{ij} = \sum_k\lambda_k
\frac{\partial X_k(t_f)}{\partial A_{ij}}
\end{equation}
In the decoupled direct method, \Eq{ddmeq} is solved for $\partial X_k/\partial A_{ij}$
using a method which exploits the sparseness of $\partial A/\partial A_{ij}=\delta_{ij}$ in the present context.
However, it is also possible to express $S_{DM}$ in terms of the matrix Fr\'{e}chet derivative
as explained in the Appendix, viz.
\begin{equation} \label{eq:SFDM}
S_{FDM}^{ij}(t_f) = t_f \sum_k\lambda_k
L_{\exp}(t_f \mathsf{A}, \mathsf{E}_{ij})X(0)
\end{equation}
where $L_{\exp}$ is the matrix Fr\'{e}chet derivative as defined in the Appendix
where $\mathsf{E}_{ij}$ is also defined.
\Eq{SFDM} defines the Fr\'{e}chet direct method. Similarly to the PBM, this
technique required modification of \textsc{Fispact-II} to output the
matrix~$\mathsf{A}$ in a format suitable for input to MATLAB~\cite{matlab}.%$^{TM}$.

\subsection{Pearson Derived Method}\label{sec:PRS}
%The sensitivity of dependent quantity $Y_j$ on independent variable $X_i$ is
The Pearson technique for ranking sensitivities starts with the definition of
the Pearson product-moment correlation coefficient for a set of $N_s$~samples
$\{(A_s,Q_s):s=1,\ldots,N_s\}$, viz.
\begin{equation} \label{eq:pearson}
r = \frac{
\sum_s (Q_s - \bar{Q}) (A_s-\bar{A})
}{(N_s-1)
\Delta Q \Delta A
}
\end{equation}
where the suffix~$ij$ on $r$~and~$A$ is to be understood, overbar denotes
average and $\Delta$ denotes the standard deviation of the distribution so that
for example
\begin{eqnarray}
\label{eq:16}
\bar{Q} &=& \frac{1}{N_s}\sum_{s=1}^{N_s} Q_s\\
\label{eq:17}
\Delta Q &=& \sqrt{\frac{1}{(N_s-1)}\sum_{s=1}^{N_s} [(Q_s)^2-\bar{Q}^2]}
\end{eqnarray}
The coefficient $r_{ij}$ is by definition always less than or equal to one,
and a magnitude of~$r$ close to one
indicates strong linear correlation.

However, it is the proportionality constant corresponding to~$\partial Q/\partial A_{ij}$
that is of initial interest. It is worth cautioning that although the definition
implicitly implies a linear relation, there is no guarantee of this. However, in order to
proceed, linearity will be assumed, viz.
\begin{equation} \label{eq:19}
Q-\bar{Q} = \tilde{r}(A-\bar{A})
\end{equation}
and substituting in \Eq{pearson}, it follows that
%The best-fit line relating $Q$ to $A$ is given by
\begin{equation} \label{eq:prtwid}
S_{PRD}=\tilde{r}_{ij}=r_{ij}\left(\frac{\Delta Q}{\Delta A}\right)
\end{equation}
It follows that the output of the Monte-Carlo sensitivity calculations
may be used to rank the different interactions by computing~$r_{ij}/\Delta A$
(note that $\Delta Q$ is the same for all the $A_{ij}$~variations in the
standard approach described in ref~\cite{Ar14Sens}). %Arter and Morgan(2014)

The calculation
of the Pearson coefficient~$r$ is well-known to be sensitive to round-off error.
To avoid modifying the software, the coefficient is computed using output
values from \textsc{Fispact-II} given
only to $6$~significant figures by default.
This accuracy is the maximum that can be expected from the numerical integration
of the rate equation which is is constrained to an accuracy of one part in a million.
It was found that splitting the separate contributions of~$A_s$ and~$\bar{A}$ to
\Eq{pearson} led to unacceptable cancellation due to round-off effects
(although it was verified that round-off was not a similar issue for $Q_s$ and  $\bar{Q}$).

\subsection{Comments upon the Different Metrics}\label{sec:comment}
The main distinction between the PBM and the other two measures is that
the pathways-based method is `global', capturing the whole variation of
the inventory as parameters are varied, although having the disadvantage
that it cannot measure sensitivity to diagonal entries of~$\sf{A}$.
The other two techniques are more local,
indeed the DM returns directly only a coefficient at the mean of the
distribution of~$Q$. The Pearson method is somewhere in-between, using global
variations, but making a local linear assumption about the mean. This
complicates the comparison in the next \Sec{calculations}.

The principal comment to be made concerning the comparison is that,
corresponding to the lack of sensitivity to element~$A_{ij}$ when it is zero
due an absence of interaction between nuclides~$i$ and~$j$, a large
sensitivity in the local sense may be inconsequential for the total activity~$Q$
if the corresponding~$A_{ij}$ is relatively very small.
However, the two more local estimates (\Eq{SDM} and
\Eq{prtwid}) for $\partial Q/\partial A_{ij}$ should be directly comparable.

Main interest attaches to global measures such as~$S_{PBM}$. The FDM
approach may be used to produce an equivalent ranking by scaling
by the estimated error in the coefficient, viz.
\begin{equation} \label{eq:SFDS}
S_{FDS}^{ij} = S_{FDM}^{ij} \cdot \left(\frac{\epsilon_{ij}}{100}\right)\cdot \bar{A}_{ij}
\end{equation}
where~$\epsilon_{ij}$ is the percentage error in the distribution
of the coefficient~$A_{ij}$.
\textsc{Fispact-II} returns both ~$\epsilon_{ij}$
and~$\bar{A}_{ij}$ by combining the uncertainties in the reaction coefficients corresponding
to~$A_{ij}$.

From \Eq{prtwid}, a ranking based on the Pearson coefficient~$r$ should
also be comparable to $S_{PBM}$, if it is scaled similarly, viz.
\begin{equation} \label{eq:SPRS}
S_{PRS}^{ij} = r_{ij} \cdot \left(\frac{\epsilon_{ij}}{100}\right)\cdot
\left(\frac{\bar{A}_{ij}}{\Delta A_{ij}}\right)
\end{equation}
In practice it is found that $S_{PRS}^{ij} \approx r_{ij}$.

Note that for interactions for which
no uncertainty information is available, a Pearson coefficient cannot be computed,
nor is $S_{FDS}$ useful. The coefficient $S_{PBM}$ may be non-zero, but this relies
on the interaction's lying on a pathway important for other reasons. Interactions
without accompanying uncertainty information will therefore largely have to be ignored in
this work.

\section{Sensitivity Calculations}\label{sec:calculations}
\subsection{Details of Cases}\label{sec:details}
The test cases are taken from ref~\cite{Ar14Sens} %Arter and Morgan(2014)
and involve several different nuclide mixtures designed to 
be indicative of a wide range of activation problems,
see \Table{mix}. As indicated, all but one of the mixtures
consisted of~$1$\,kg of material subject to a neutron flux of $10^{15}$\,$cm^{-2}s^{-1}$,
for a year, without any cooling period.

The mixtures are used in six test cases, with the Alloy case extended to
include a cooling phase. 
Each test case is run using the full TENDL~2013 data from the
EASY-II database~\cite{easy} with pathways analysis
to identify the important reactions, the numbers of
which are listed in \Table{stats}\@. 
As in ref~\cite{Ar14Sens}, %Arter and Morgan(2014)
Monte-Carlo solution of the reduced 
problem, investigating the distributions of the important reaction rates specified
in the newer database, was
then performed in the sequence of increasing sample size per reaction,
$N_x=10,\;40,\;160,\;\ldots$ up to the maximum value specified in the table.
Indications from ref~\cite{Ar14Sens} %Arter and Morgan(2014)
and work which may be published elsewhere indicate that the pathways-reduced results
agree to at least two (and often three) significant figures with those obtained
by sampling the full problem, at less than a thousandth of the computational cost.
As might be expected from the large maximum number of samples~$N_s$ employed, the
distributions of reaction rates actually sampled usually agree in the mean 
to $4$~significant figures with the nominal database values.

\begin{table*}
\caption{Test cases. Each consists of
numbers of atoms of the listed elements with their natural abundances of nuclides,
given as percentages by mass of the whole.}
\label{tab:mix}
\begin{center}
\begin{tabular}{|p{1.5cm}|p{6.5cm}|c|c|c|c|}
\hline
Test & Constituents of Mixture & Sample & Irradiation & Cooling & Neutron flux\\
Label & & Mass & Period & Period & cm$^{-2}$s$^{-1}$ \\
\hline
Alloy & Fe $40.0$ : Ni $20.0$ : Cr $20.0$ : Mn $20.0$ & $1$\,kg & $1$\,yr & 0 & $10^{15}$\\
Alloy+c & Fe $40.0$ : Ni $20.0$ : Cr $20.0$ : Mn $20.0$ & $1$\,kg & $1$\,yr & $1$\,yr & $10^{15}$\\
Fe & Fe  & $1$\,kg & $2.5$\,yr & 0 & $10^{15}$ \\
%Fiss & U235 $3.7$ : U238 $96.3$ & $8.7$\,g & $3$\,mo & 0 & $2.59032 \times 10^{14}$ \\
LiMix & Li $40.0$ : Be $30.0$ : O $30.0$ & $1$\,kg & $1$\,yr & 0 & $10^{15}$\\
WMix & W $20.0$ : Re $20.0$ : Ir $20.0$ : Bi $20.0$ : Pb $20.0$ & $1$\,kg & $1$\,yr & 0 & $10^{15}$\\
Y2O3 & Y  $78.74$ : O  $21.26$ & $1$\,g & $300$\,s & 0 & $1.116 \times 10^{10}$\\
\hline
\end{tabular}
\end{center}
\end{table*}
\begin{table*}
\caption{Test cases statistics. 
Monte-Carlo sampling by \textsc{Fispact-II} has a sample
size determined by the number of reactions examined.
}
\label{tab:stats}
\begin{center}
\begin{tabular}{|p{1.5cm}|c|c|c|c|}
\hline
Test & $I$,\,Reactions  & Matrix & Max. $N_x$,\,Samples & $N_s$,\,Total  \\
Label & Examined  & $\mathsf{A}$\,Size & per Reaction & Sample  \\
\hline
Alloy  & $84$ & $51$ & $640$ & $53\,760$  \\
Alloy+c  & $50$  & $38$ & $640$ & $32\,000$  \\
Fe  & $27$ & $24$  & $640$ & $17\,280$  \\
LiMix  & $17$ & $21$ & $640$ & $10\,880$  \\
WMix  & $71$ & $63$ & $640$ & $45\,440$  \\
Y2O3  & $13$ &  $16$ & $2\,560$ & $33\,280$  \\
\hline
\end{tabular}
\end{center}
\end{table*}

\clearpage
\subsection{Results}\label{sec:results}
This section presents results for each of the test cases in turn, 
in the alphabetic order specified in \Table{mix}. Attention is drawn to the fact
that the Y2O3 case is the simplest in terms of pathways, and contains
extra explanation.

For each test case there is a table of sensitivity rankings ordered by Fr\'{e}chet derivative
amplitude and a graph of rankings ordered by $S_{PBM}$.
The table enables a larger range of interactions to be compared, since the
graphs become hard to interpret once the number of plotted interactions
exceeds about ten.
Note the convention (except for the Y2O3 case) that all three methods must
provide a ranking for the comparison to be plotted. So in the figures
the ten highest-ranked cases plotted may  include reactions significantly
smaller in importance than the tenth.

A general feature of all graphs comparing rankings by the different techniques
is the symmetry about the mid-line labelled~$PBM$. The appearance of ``$V$" and
``$\Lambda$" patterns indicates that although the more local measures may not agree
with $S_{PBM}$, they do themselves correlate well.
%Unfortunately, the absence from the plots of a small fraction of interactions highly ranked
%by the PBM is usually because they have not been ranked by the FDM. This fraction
%reaches~$80\,\%$ in \Sec{Y2O3}, hence the convention was adopted there to plot
%all the first ten as ranked by PBM, assigning zero to interactions not ranked by
%the other methods.

For three of the test cases, Alloy+c in \Sec{cSt}, WMix in \Sec{WMix} and
Y2O3 in \Sec{Y2O3}, further
results of analysis are presented to help understand the effect of sampling
and round-off on the calculation of~$S_{PRS}$. In addition, a table of sensitivity
rankings ordered by $S_{PBM}$ and a graph of rankings ordered by Fr\'{e}chet derivative
also appear in these two sections. (This information is omitted from the other
four sections \Sec{Steel}, \Sec{Fe}, \Sec{LiMix} and \Sec{Y2O3} to save space.)
\subsubsection{Alloy}\label{sec:Steel}
\begin{figure}[htbp]
\centerline{\rotatebox{270}{\includegraphics[width=9.0cm]{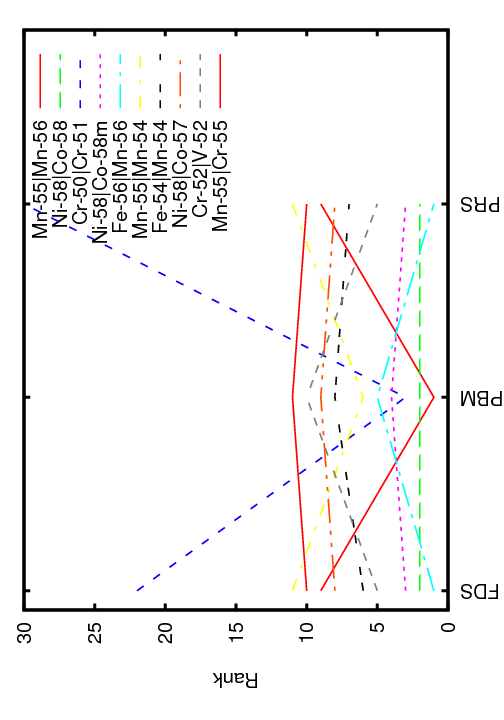}}}
\caption{\label{fig:cfSteel_rrank}
Comparison of the Alloy test case results, showing the first ten interactions
according to the Pathways Based Metric~$S_{PBM}$, ranked accordingly.}
\end{figure}

See \Table{Steel} and \Fig{cfSteel_rrank}.
\begin{table}
\caption{Alloy case.
 Rankings for different methods.
}
\label{tab:Steel}
\begin{tabular}{|p{2.0cm}|p{2.0cm}|c|c|c|}
\hline
& & \multicolumn{3}{|c|}{Sensitivity} \\
\hline
 Parent & Child & FDS & PBM & PRS \\
\hline
 Fe-56 & Mn-56 & $1$ & $5$ & $1$ \\
 Ni-58 & Ni-57 & $21$ & $37$ & $57$ \\
 Mn-55 & Cr-55 & $10$ & $11$ & $10$ \\
 Mn-55 & V-52 & $12$ & $18$ & $12$ \\
 Mn-55 & Mn-56 & $9$ & $1$ & $9$ \\
 Cr-52 & Cr-51 & $17$ & $19$ & $24$ \\
 Cr-52 & V-52 & $5$ & $10$ & $5$ \\
 Ni-58 & Co-58m & $3$ & $4$ & $3$ \\
 Mn-55 & Mn-54 & $11$ & $6$ & $11$ \\
 Fe-56 & Fe-55 & $14$ & $15$ & $15$ \\
 Ni-58 & Co-57 & $8$ & $9$ & $8$ \\
 Ni-60 & Co-60m & $4$ & $13$ & $4$ \\
 Ni-58 & Co-58 & $2$ & $2$ & $2$ \\
 Ni-58 & Fe-55 & $7$ & $12$ & $6$ \\
 Fe-54 & Cr-51 & $13$ & $20$ & $13$ \\
 Cr-53 & V-52 & $37$ & $0$ & $0$ \\
 Cr-53 & V-53 & $15$ & $25$ & $18$ \\
 Fe-54 & Mn-54 & $6$ & $8$ & $7$ \\
 Cr-50 & Cr-51 & $22$ & $3$ & $30$ \\
 Fe-57 & Mn-56 & $36$ & $0$ & $0$ \\
 Fe-57 & Mn-57 & $18$ & $26$ & $33$ \\
 Ni-62 & Fe-59 & $19$ & $35$ & $71$ \\
 Ni-62 & Co-62m & $30$ & $50$ & $56$ \\
 Ni-62 & Co-61 & $39$ & $79$ & $66$ \\
 Ni-62 & Co-62 & $28$ & $46$ & $47$ \\
 Ni-60 & Co-60 & $16$ & $28$ & $14$ \\
 Cr-54 & Cr-55 & $24$ & $22$ & $67$ \\
 Cr-54 & Ti-51 & $27$ & $49$ & $26$ \\
 Cr-54 & V-54 & $25$ & $52$ & $50$ \\
 Fe-54 & Fe-55 & $26$ & $14$ & $48$ \\
%X  Fe-56 & Mn-55 & $31$ & $0$ & $0$ \\
%X  Cr-50 & V-49 & $23$ & $27$ & $70$ \\
%X  Ni-64 & Fe-61 & $33$ & $55$ & $74$ \\
%X  Ni-61 & Co-60m & $38$ & $0$ & $0$ \\
%X  Ni-61 & Co-61 & $20$ & $32$ & $78$ \\
%X  Ni-60 & Co-59 & $29$ & $41$ & $36$ \\
%X  Ni-64 & Ni-65 & $32$ & $21$ & $38$ \\
%X  Fe-58 & Cr-55 & $34$ & $64$ & $83$ \\
%X  Fe-58 & Mn-58 & $35$ & $66$ & $68$ \\
\hline
\end{tabular}
\end{table}

%X \begin{figure}[htbp]
%X \centerline{\rotatebox{270}{\includegraphics[width=9.0cm]{cfSteel_drank}}}
%X \caption{\label{fig:cfSteel_drank}
%X Comparison of the Alloy test case results, showing the first ten interactions
%X by magnitude of Fr\'{e}chet derivative. The labels are ordered according to
%X Fr\'{e}chet derivative size, so that the top interaction is the most sensitive.} 
%X \end{figure}
%X \input{rSteel}
\clearpage
\subsubsection{Alloy+c}\label{sec:cSt}
\begin{figure}[htbp]
\centerline{\rotatebox{270}{\includegraphics[width=9.0cm]{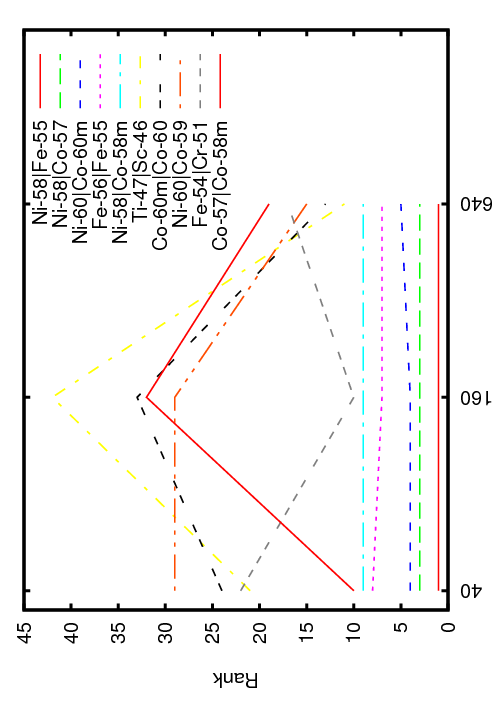}}}
\caption{\label{fig:cfcSt_n3rank}
Comparison of the Alloy+c test case results, showing the first ten 
odd-numbered interactions
according to the scaled Pearson technique value~$S_{PRS}$, for a
Monte-Carlo sample size of~$N_x=640$ per reaction, as $N_x$ is increased.}
\end{figure}

\begin{figure}[htbp]
\centerline{\rotatebox{270}{\includegraphics[width=9.0cm]{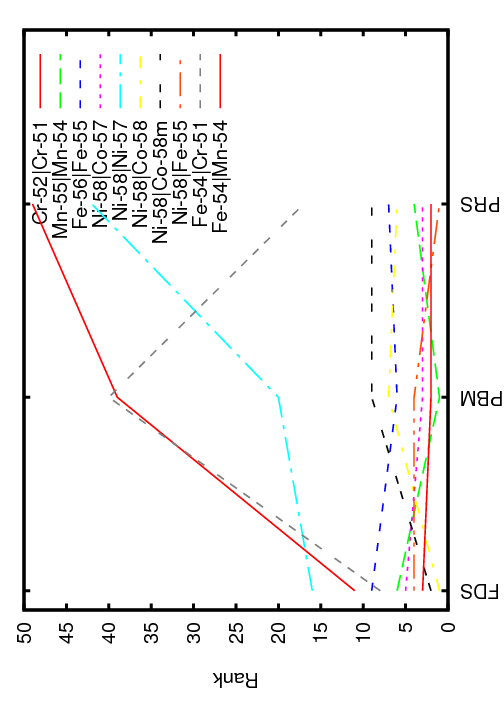}}}
\caption{\label{fig:cfcSt_drank}
Comparison of the Alloy+c test case results, showing the first ten interactions
by magnitude of Fr\'{e}chet derivative. The labels are ordered according to
Fr\'{e}chet derivative size, so that the top interaction is the most sensitive.} 
\end{figure}

\begin{figure}[htbp]
\centerline{\rotatebox{270}{\includegraphics[width=9.0cm]{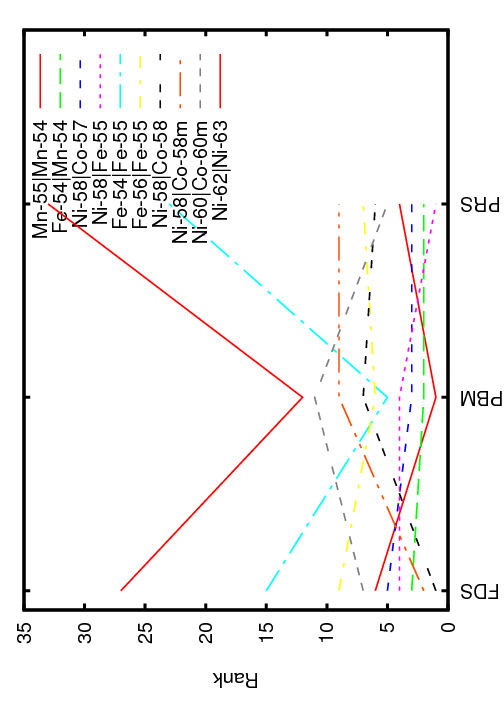}}}
\caption{\label{fig:cfcSt_rrank}
Comparison of the Alloy+c test case results, showing the first ten interactions
according to the Pathways Based Metric~$S_{PBM}$ for which comparison is possible.}
\end{figure}

See \Table{PcSt}, \Table{dcSt}, \Table{rcSt}, \Fig{cfcSt_n3rank},  \Fig{cfcSt_drank} and  \Fig{cfcSt_rrank}.
%The  FDM analysis resulted in a warning that the matrix~$\sf{A}$ was ill-conditioned.

\Table{PcSt} suggests that once the Pearson correlation becomes below~$0.1$
it becomes inaccurate. \Fig{cfcSt_n3rank} shows that the smaller Pearson coefficients
vary erratically with sampling, from which it is inferred that
round-off effects have become important.

As indicated in \Table{mix} this case involves both an irradiation phase and a 
cooling period.
Care is required in comparing the FDM approach in this instance, for the method
uses only the matrix for the cooling phase, whereas the other analyses are of the
entire history. Although there is still reasonably good correlation between
$P_{FDS}$ and $P_{PRS}$, it is not as good in the other test cases.
\begin{table}
\caption{Alloy+c case.
 Values of absolute Pearson correlation coefficient~$|r|$
as Monte Carlo sample size increases with~$N_x$.
}
\label{tab:PcSt}
\begin{tabular}{|p{2.0cm}|p{2.0cm}|c|c|c|}
\hline
& & \multicolumn{3}{|c|}{Absolute Pearson} \\
\hline
 Parent & Child & $40$ & $160$ & $640$ \\
\hline
Ni-58 & Fe-55 & $  0.81342$ & $  0.79097$ & $  0.79709$ \\
Fe-54 & Mn-54 & $  0.48716$ & $  0.48798$ & $  0.48761$ \\
Ni-58 & Co-57 & $  0.24152$ & $  0.23469$ & $  0.23442$ \\
Mn-55 & Mn-54 & $  0.17194$ & $  0.16534$ & $  0.16563$ \\
Ni-60 & Co-60m & $  0.16462$ & $  0.15008$ & $  0.13389$ \\
Ni-58 & Co-58 & $  0.13754$ & $  0.12544$ & $  0.11864$ \\
Fe-56 & Fe-55 & $  9.410\times 10^{-2}$ & $  7.202\times 10^{-2}$ & $  9.237\times 10^{-2}$ \\
Ni-60 & Co-60 & $  0.10390$ & $  6.218\times 10^{-2}$ & $  5.669\times 10^{-2}$ \\
Ni-58 & Co-58m & $  6.996\times 10^{-2}$ & $  5.098\times 10^{-2}$ & $  4.101\times 10^{-2}$ \\
Ti-46 & Sc-46m & $  3.614\times 10^{-2}$ & $  1.570\times 10^{-3}$ & $  1.575\times 10^{-2}$ \\
Ti-47 & Sc-46 & $  2.538\times 10^{-2}$ & $  2.806\times 10^{-3}$ & $  1.209\times 10^{-2}$ \\
V-49 & Sc-46 & $  4.678\times 10^{-3}$ & $  1.198\times 10^{-2}$ & $  1.144\times 10^{-2}$ \\
Co-60m & Co-60 & $  1.923\times 10^{-2}$ & $  9.042\times 10^{-3}$ & $  8.303\times 10^{-3}$ \\
Cr-50 & V-50 & $  9.279\times 10^{-3}$ & $  1.598\times 10^{-2}$ & $  8.073\times 10^{-3}$ \\
Ni-60 & Co-59 & $  1.648\times 10^{-2}$ & $  1.167\times 10^{-2}$ & $  7.639\times 10^{-3}$ \\
Ni-57 & Co-57 & $  1.665\times 10^{-2}$ & $  1.499\times 10^{-2}$ & $  7.389\times 10^{-3}$ \\
Co-57 & Co-58 & $  2.483\times 10^{-3}$ & $  1.572\times 10^{-2}$ & $  6.780\times 10^{-3}$ \\
Fe-54 & Cr-51 & $  2.407\times 10^{-2}$ & $  2.252\times 10^{-2}$ & $  6.773\times 10^{-3}$ \\
Co-57 & Co-58m & $  3.906\times 10^{-2}$ & $  9.176\times 10^{-3}$ & $  6.739\times 10^{-3}$ \\
Ti-47 & Ti-46 & $  2.828\times 10^{-2}$ & $  1.220\times 10^{-2}$ & $  5.516\times 10^{-3}$ \\
Co-59 & Fe-59 & $  8.709\times 10^{-3}$ & $  1.451\times 10^{-2}$ & $  5.499\times 10^{-3}$ \\
Fe-58 & Fe-59 & $  2.626\times 10^{-2}$ & $  1.466\times 10^{-2}$ & $  5.289\times 10^{-3}$ \\
Fe-54 & Fe-55 & $  2.698\times 10^{-2}$ & $  7.802\times 10^{-3}$ & $  5.261\times 10^{-3}$ \\
Co-58 & Fe-58 & $  2.502\times 10^{-3}$ & $  1.069\times 10^{-3}$ & $  5.199\times 10^{-3}$ \\
Ti-47 & Sc-46m & $  4.156\times 10^{-3}$ & $  1.867\times 10^{-2}$ & $  5.039\times 10^{-3}$ \\
Cr-50 & Cr-51 & $  1.015\times 10^{-2}$ & $  2.073\times 10^{-3}$ & $  4.815\times 10^{-3}$ \\
Co-59 & Co-60 & $  1.789\times 10^{-2}$ & $  1.673\times 10^{-2}$ & $  4.743\times 10^{-3}$ \\
Ni-58 & Ni-59 & $  2.777\times 10^{-3}$ & $  1.273\times 10^{-2}$ & $  4.581\times 10^{-3}$ \\
Mn-55 & Mn-56 & $  1.473\times 10^{-2}$ & $  3.741\times 10^{-3}$ & $  4.456\times 10^{-3}$ \\
Co-58 & Co-59 & $  5.478\times 10^{-3}$ & $  6.033\times 10^{-4}$ & $  4.220\times 10^{-3}$ \\
\hline
\end{tabular}
\end{table}

\begin{table}
\caption{Alloy+c case.
 Rankings for different methods.
}
\label{tab:dcSt}
\begin{tabular}{|p{2.0cm}|p{2.0cm}|c|c|c|}
\hline
& & \multicolumn{3}{|c|}{Sensitivity} \\
\hline
 Parent & Child & FDS & PBM & PRS \\
\hline
 Cr-52 & Cr-51 & $11$ & $39$ & $49$ \\
 Mn-55 & Mn-54 & $6$ & $1$ & $4$ \\
 Fe-56 & Fe-55 & $9$ & $6$ & $7$ \\
 Ni-58 & Co-57 & $5$ & $3$ & $3$ \\
 Ni-58 & Ni-57 & $16$ & $20$ & $42$ \\
 Ni-58 & Co-58 & $1$ & $7$ & $6$ \\
 Ni-58 & Co-58m & $2$ & $9$ & $9$ \\
 Ni-58 & Fe-55 & $4$ & $4$ & $1$ \\
 Fe-54 & Cr-51 & $8$ & $40$ & $17$ \\
 Fe-54 & Mn-54 & $3$ & $2$ & $2$ \\
 Cr-50 & Cr-51 & $13$ & $26$ & $26$ \\
 Ni-62 & Fe-59 & $12$ & $37$ & $44$ \\
 Ni-60 & Co-60 & $10$ & $14$ & $8$ \\
 Ni-60 & Co-60m & $7$ & $11$ & $5$ \\
 Fe-54 & Fe-55 & $15$ & $5$ & $23$ \\
 Cr-50 & V-49 & $14$ & $16$ & $35$ \\
 Fe-58 & Fe-59 & $18$ & $27$ & $22$ \\
 Fe-54 & Mn-53 & $17$ & $18$ & $30$ \\
 Ni-60 & Co-59 & $22$ & $28$ & $15$ \\
 Ni-59 & Co-58 & $21$ & $0$ & $0$ \\
 Ni-58 & Fe-54 & $31$ & $0$ & $0$ \\
 Ni-59 & Co-58m & $23$ & $0$ & $0$ \\
 Co-59 & Fe-59 & $19$ & $42$ & $20$ \\
 Co-59 & Co-58 & $26$ & $0$ & $0$ \\
 Fe-56 & Mn-55 & $30$ & $0$ & $0$ \\
 Co-59 & Co-58m & $25$ & $0$ & $0$ \\
 Ni-59 & Fe-55 & $32$ & $0$ & $0$ \\
 Ni-62 & Ni-63 & $27$ & $12$ & $33$ \\
 Fe-55 & Mn-54 & $28$ & $0$ & $0$ \\
 Co-58 & Co-57 & $24$ & $30$ & $34$ \\
 Co-57 & Co-56 & $29$ & $44$ & $37$ \\
 Co-58 & Mn-54 & $33$ & $0$ & $0$ \\
 Co-58 & Co-58m & $20$ & $62$ & $0$ \\
\hline
\end{tabular}
\end{table}

\begin{table}
\caption{Alloy+c case.
 Rankings for different methods.
}
\label{tab:rcSt}
\begin{tabular}{|p{2.0cm}|p{2.0cm}|c|c|c|}
\hline
& & \multicolumn{3}{|c|}{Sensitivity} \\
\hline
 Parent & Child & FDS & PBM & PRS \\
\hline
 Mn-55 & Mn-54 & $6$ & $1$ & $4$ \\
 Fe-54 & Mn-54 & $3$ & $2$ & $2$ \\
 Ni-58 & Co-57 & $5$ & $3$ & $3$ \\
 Ni-58 & Fe-55 & $4$ & $4$ & $1$ \\
 Fe-54 & Fe-55 & $15$ & $5$ & $23$ \\
 Fe-56 & Fe-55 & $9$ & $6$ & $7$ \\
 Ni-58 & Co-58 & $1$ & $7$ & $6$ \\
 Co-60m & Co-60 & $0$ & $8$ & $13$ \\
 Ni-58 & Co-58m & $2$ & $9$ & $9$ \\
 Co-58m & Co-58 & $0$ & $10$ & $32$ \\
 Ni-60 & Co-60m & $7$ & $11$ & $5$ \\
 Ni-62 & Ni-63 & $27$ & $12$ & $33$ \\
 Co-58 & Co-59 & $0$ & $13$ & $31$ \\
 Ni-60 & Co-60 & $10$ & $14$ & $8$ \\
 Co-59 & Co-60m & $0$ & $15$ & $43$ \\
 Cr-50 & V-49 & $14$ & $16$ & $35$ \\
 Co-59 & Co-60 & $0$ & $17$ & $27$ \\
 Fe-54 & Mn-53 & $17$ & $18$ & $30$ \\
 Mn-53 & Mn-54 & $0$ & $19$ & $46$ \\
 Ni-57 & Co-57 & $0$ & $21$ & $16$ \\
 Ni-58 & Ni-57 & $16$ & $20$ & $42$ \\
 Mn-55 & Mn-56 & $0$ & $23$ & $29$ \\
 Mn-56 & Fe-56 & $0$ & $22$ & $0$ \\
 Co-58m & Co-59 & $0$ & $24$ & $36$ \\
 Co-57 & Co-58 & $0$ & $25$ & $18$ \\
 Cr-50 & Cr-51 & $13$ & $26$ & $26$ \\
 Fe-58 & Fe-59 & $18$ & $27$ & $22$ \\
 Ni-60 & Co-59 & $22$ & $28$ & $15$ \\
 Ni-64 & Ni-63 & $0$ & $29$ & $39$ \\
 Co-58 & Co-57 & $24$ & $30$ & $34$ \\
%X  Co-57 & Co-58m & $0$ & $31$ & $19$ \\
%X  Ni-58 & Ni-59 & $0$ & $32$ & $28$ \\
%X  Fe-59 & Co-59 & $0$ & $33$ & $0$ \\
%X  Mn-55 & H-3 & $0$ & $34$ & $0$ \\
%X  V-50 & V-49 & $0$ & $35$ & $48$ \\
%X  Cr-50 & V-50 & $0$ & $36$ & $14$ \\
%X  Ni-62 & Fe-59 & $12$ & $37$ & $44$ \\
%X  Co-58 & Fe-58 & $0$ & $38$ & $24$ \\
%X  Cr-52 & Cr-51 & $11$ & $39$ & $49$ \\
%X  Fe-54 & Cr-51 & $8$ & $40$ & $17$ \\
%X  Fe-57 & Fe-58 & $0$ & $41$ & $40$ \\
%X  Co-59 & Fe-59 & $19$ & $42$ & $20$ \\
%X  Ni-60 & Ni-59 & $0$ & $43$ & $41$ \\
%X  Co-57 & Co-56 & $29$ & $44$ & $37$ \\
%X  Cr-53 & H-3 & $0$ & $45$ & $0$ \\
%X  Fe-57 & H-3 & $0$ & $46$ & $0$ \\
%X  He-3 & H-3 & $0$ & $48$ & $0$ \\
%X  Ni-58 & He-3 & $0$ & $47$ & $0$ \\
%X  V-49 & Sc-46 & $0$ & $49$ & $12$ \\
%X  Sc-46m & Sc-46 & $0$ & $50$ & $45$ \\
%X  V-49 & Sc-46m & $0$ & $51$ & $47$ \\
%X  Cr-50 & Ti-47 & $0$ & $52$ & $50$ \\
%X  Ti-47 & Sc-46 & $0$ & $53$ & $11$ \\
%X  Ti-47 & Sc-46m & $0$ & $54$ & $25$ \\
%X  Mn-54 & Mn-53 & $0$ & $55$ & $0$ \\
%X  Ti-47 & Ti-46 & $0$ & $56$ & $21$ \\
%X  Ti-46 & Sc-46 & $0$ & $57$ & $38$ \\
%X  Ti-46 & Sc-46m & $0$ & $58$ & $10$ \\
%X  Sc-46 & Sc-46m & $0$ & $59$ & $0$ \\
%X  Co-60 & Co-60m & $0$ & $61$ & $0$ \\
%X  H-3 & He-3 & $0$ & $60$ & $0$ \\
%X  Co-58 & Co-58m & $20$ & $62$ & $0$ \\
\hline
\end{tabular}
\end{table}

\clearpage
\subsubsection{Fe}\label{sec:Fe}
\begin{figure}[htbp]
\centerline{\rotatebox{270}{\includegraphics[width=9.0cm]{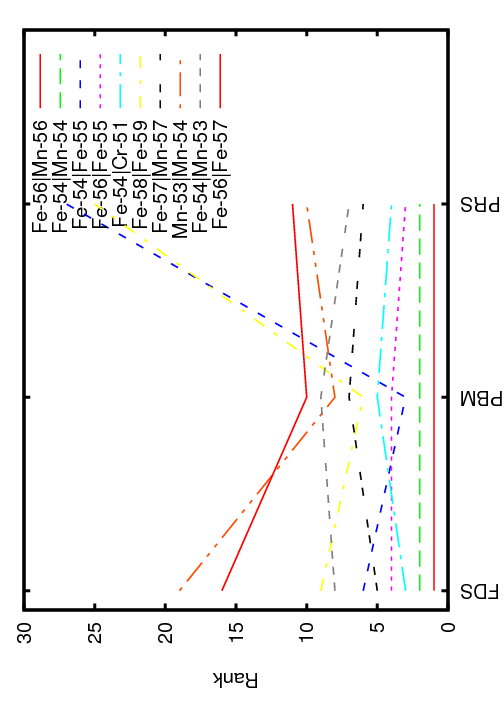}}}
\caption{\label{fig:cfFe_rrank}
Comparison of the Fe test case results, showing the first ten interactions
according to the Pathways Based Metric~$S_{PBM}$, ranked accordingly.}
\end{figure}

See \Table{Fe} and \Fig{cfFe_rrank}.
\begin{table}
\caption{Fe case.
 Rankings for different methods.
}
\label{tab:Fe}
\begin{tabular}{|p{2.0cm}|p{2.0cm}|c|c|c|}
\hline
& & \multicolumn{3}{|c|}{Sensitivity} \\
\hline
 Parent & Child & FDS & PBM & PRS \\
\hline
 Fe-56 & Mn-56 & $1$ & $1$ & $1$ \\
 Fe-56 & Fe-55 & $4$ & $4$ & $3$ \\
 Fe-54 & Cr-51 & $3$ & $5$ & $4$ \\
 Fe-54 & Mn-54 & $2$ & $2$ & $2$ \\
 Fe-57 & Mn-56 & $13$ & $0$ & $0$ \\
 Fe-57 & Mn-57 & $5$ & $7$ & $6$ \\
 Fe-54 & Fe-55 & $6$ & $3$ & $27$ \\
 Fe-56 & Mn-55 & $7$ & $27$ & $18$ \\
 Fe-58 & Fe-59 & $9$ & $6$ & $25$ \\
 Fe-58 & Cr-55 & $10$ & $19$ & $8$ \\
 Fe-58 & Mn-58m & $11$ & $21$ & $19$ \\
 Fe-58 & Mn-58 & $12$ & $22$ & $22$ \\
 Fe-54 & Mn-53 & $8$ & $9$ & $7$ \\
 Fe-56 & Fe-57 & $16$ & $10$ & $11$ \\
 Fe-55 & Mn-54 & $14$ & $0$ & $0$ \\
 Cr-54 & Cr-55 & $20$ & $15$ & $12$ \\
 Cr-54 & V-54 & $21$ & $26$ & $5$ \\
 Mn-55 & Cr-55 & $18$ & $24$ & $26$ \\
 Mn-55 & V-52 & $22$ & $25$ & $16$ \\
 Mn-55 & Mn-56 & $15$ & $0$ & $0$ \\
 Fe-57 & Fe-58 & $25$ & $13$ & $24$ \\
 Mn-55 & Mn-54 & $24$ & $0$ & $0$ \\
 Fe-57 & Cr-54 & $27$ & $28$ & $14$ \\
 V-51 & V-52 & $32$ & $16$ & $23$ \\
 Mn-53 & Mn-54 & $19$ & $8$ & $10$ \\
 Fe-57 & Fe-56 & $23$ & $0$ & $0$ \\
 Co-59 & Co-60m & $30$ & $12$ & $21$ \\
 Co-59 & Mn-56 & $31$ & $0$ & $0$ \\
 Co-59 & Fe-59 & $28$ & $0$ & $0$ \\
 Fe-55 & Mn-55 & $17$ & $23$ & $17$ \\
%X  Mn-54 & Mn-55 & $26$ & $0$ & $0$ \\
%X  Mn-54 & Mn-53 & $29$ & $0$ & $0$ \\
\hline
\end{tabular}
\end{table}

%X \begin{figure}[htbp]
%X \centerline{\rotatebox{270}{\includegraphics[width=9.0cm]{cfFe_drank}}}
%X \caption{\label{fig:cfFe_drank}
%X Comparison of the Fe test case results, showing the first ten interactions
%X by magnitude of Fr\'{e}chet derivative. The labels are ordered according to
%X Fr\'{e}chet derivative size, so that the top interaction is the most sensitive.} 
%X \end{figure}
%X \input{rFe}
\clearpage
\subsubsection{LiMix}\label{sec:LiMix}
\begin{figure}[htbp]
\centerline{\rotatebox{270}{\includegraphics[width=9.0cm]{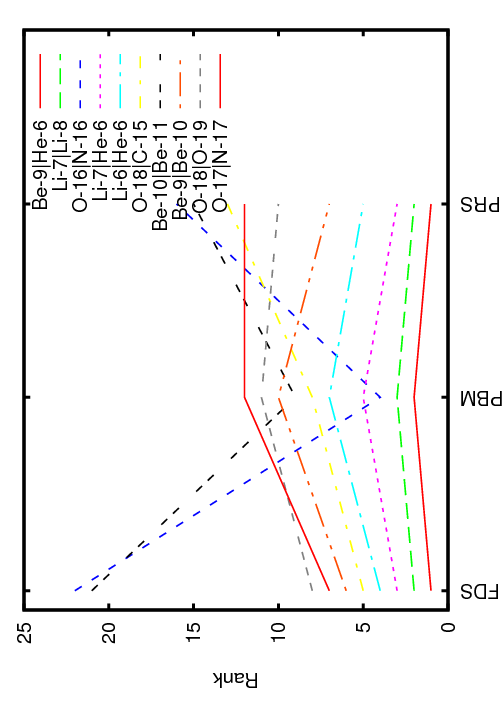}}}
\caption{\label{fig:cfLiMix_rrank}
Comparison of the LiMix test case results, showing the first ten interactions
according to the Pathways Based Metric~$S_{PBM}$ for which comparison is possible.}
\end{figure}

See \Table{LiMix} and \Fig{cfLiMix_rrank}.
%The  FDM analysis resulted in a warning that the matrix~$\sf{A}$ was ill-conditioned.
The comparison between the various metrics
in \Fig{cfLiMix_rrank} does not at first appear to be as successful as
in other cases. However the dominant interaction from
the PBM involves tritium for which uncertainty data are not accessible in the
database, hence the FDS and PRS cannot assign it a ranking and it is omitted from
the plot. Moreover all FDS rankings over~$21$ similarly correspond to zero uncertainty
and allowing for this, the comparison is as good as any reported herein.
\begin{table}
\caption{LiMix case.
 Rankings for different methods.
}
\label{tab:LiMix}
\begin{tabular}{|p{2.0cm}|p{2.0cm}|c|c|c|}
\hline
& & \multicolumn{3}{|c|}{Sensitivity} \\
\hline
 Parent & Child & FDS & PBM & PRS \\
\hline
 Li-7 & He-6 & $3$ & $5$ & $3$ \\
 Li-7 & Li-8 & $2$ & $3$ & $2$ \\
 Be-9 & He-6 & $1$ & $2$ & $1$ \\
 O-16 & N-16 & $22$ & $4$ & $16$ \\
 Li-6 & He-6 & $4$ & $7$ & $5$ \\
 Li-7 & Li-6 & $24$ & $0$ & $0$ \\
 Be-9 & Be-10 & $6$ & $10$ & $7$ \\
 O-18 & O-19 & $8$ & $11$ & $10$ \\
 O-18 & C-15 & $5$ & $8$ & $13$ \\
 O-17 & N-16 & $15$ & $0$ & $0$ \\
 O-17 & N-17 & $7$ & $12$ & $12$ \\
 O-16 & N-15 & $18$ & $14$ & $11$ \\
 Be-9 & Li-7 & $11$ & $0$ & $0$ \\
 He-3 & H-3 & $25$ & $0$ & $0$ \\
 Be-10 & He-6 & $27$ & $0$ & $0$ \\
 Be-10 & Be-11 & $21$ & $9$ & $15$ \\
 O-16 & O-17 & $14$ & $18$ & $4$ \\
 Li-6 & Li-7 & $13$ & $0$ & $0$ \\
 N-15 & N-16 & $19$ & $0$ & $0$ \\
 N-15 & C-15 & $12$ & $0$ & $0$ \\
 N-15 & B-12 & $9$ & $13$ & $6$ \\
 C-13 & Be-10 & $10$ & $16$ & $8$ \\
 O-16 & C-13 & $23$ & $15$ & $17$ \\
 C-13 & Be-9 & $26$ & $0$ & $0$ \\
 O-17 & O-18 & $20$ & $0$ & $0$ \\
 O-17 & N-15 & $17$ & $20$ & $9$ \\
 O-17 & O-16 & $16$ & $0$ & $0$ \\
\hline
\end{tabular}
\end{table}

%X \begin{figure}[htbp]
%X \centerline{\rotatebox{270}{\includegraphics[width=9.0cm]{cfLiMix_drank}}}
%X \caption{\label{fig:cfLiMix_drank}
%X Comparison of the LiMix test case results, showing the first ten interactions
%X by magnitude of Fr\'{e}chet derivative. The labels are ordered according to
%X Fr\'{e}chet derivative size, so that the top interaction is the most sensitive.} 
%X \end{figure}
%X \input{rLiMix}
\clearpage
\subsubsection{WMix}\label{sec:WMix}
\begin{figure}[htbp]
\centerline{\rotatebox{270}{\includegraphics[width=9.0cm]{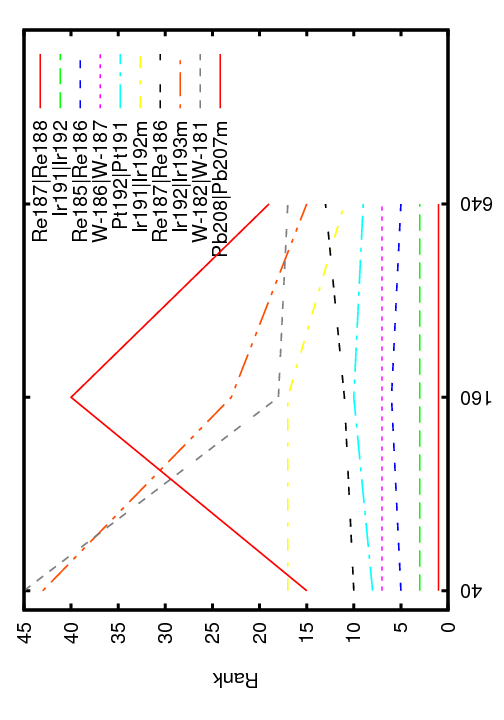}}}
\caption{\label{fig:cfWMix_n3rank}
Comparison of the WMix test case results, showing the first ten 
odd-numbered interactions
according to the scaled Pearson technique value~$S_{PRS}$, for a
Monte-Carlo sample size of~$N_x=640$ per reaction, as $N_x$ is increased.}
\end{figure}

\begin{figure}[htbp]
\centerline{\rotatebox{270}{\includegraphics[width=9.0cm]{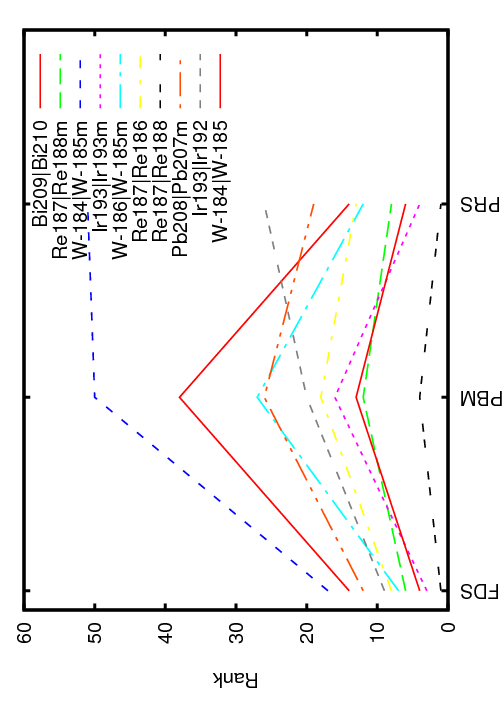}}}
\caption{\label{fig:cfWMix_drank}
Comparison of the WMix test case results, showing the first ten interactions
by magnitude of Fr\'{e}chet derivative. The labels are ordered according to
Fr\'{e}chet derivative size, so that the top interaction is the most sensitive.} 
\end{figure}

\begin{figure}[htbp]
\centerline{\rotatebox{270}{\includegraphics[width=9.0cm]{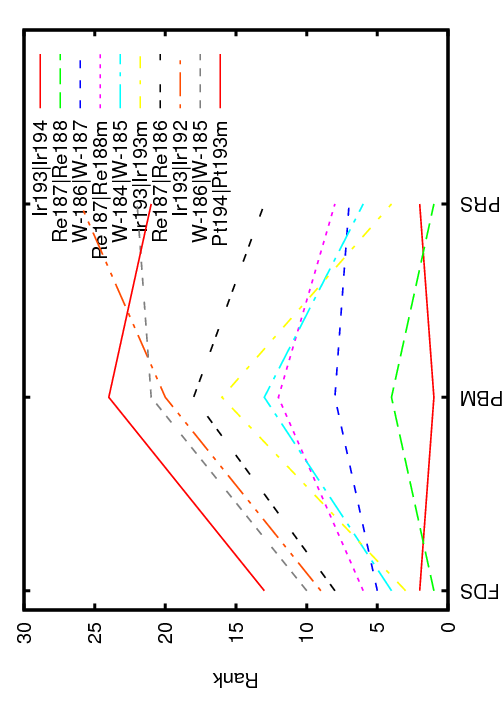}}}
\caption{\label{fig:cfWMix_rrank}
Comparison of the WMix test case results, showing the first ten interactions
according to the Pathways Based Metric~$S_{PBM}$ for which comparison is possible.}
\end{figure}

See \Table{PWMix}, , \Table{dWMix}, \Table{rWMix}, \Fig{cfWMix_n3rank},  \Fig{cfWMix_drank} and  \Fig{cfWMix_rrank}.
\Table{PWMix} suggests that once the Pearson correlation becomes below~$0.1$
it becomes inaccurate. \Fig{cfWMix_n3rank} shows that the lower rankings
in terms of sensitivity vary erratically with sampling for similar reasons
to do with round-off effects.
\begin{table}
\caption{WMix case.
 Values of absolute Pearson correlation coefficient~$|r|$
as Monte Carlo sample size increases with~$N_x$.
}
\label{tab:PWMix}
\begin{tabular}{|p{2.0cm}|p{2.0cm}|c|c|c|}
\hline
& & \multicolumn{3}{|c|}{Absolute Pearson} \\ 
\hline
 Parent & Child & $40$ & $160$ & $640$ \\ 
\hline
Re187 & Re188 & $  0.93702$ & $  0.94033$ & $  0.94091$ \\
Ir193 & Ir194 & $  0.24971$ & $  0.23407$ & $  0.24135$ \\
Ir191 & Ir192 & $  0.13306$ & $  0.13520$ & $  0.13542$ \\
Ir193 & Ir193m & $  0.11155$ & $  0.11969$ & $  0.11550$ \\
Re185 & Re186 & $  0.10156$ & $  8.413\times 10^{-2}$ & $  7.955\times 10^{-2}$ \\
W-184 & W-185 & $  6.398\times 10^{-2}$ & $  9.215\times 10^{-2}$ & $  7.856\times 10^{-2}$ \\
W-186 & W-187 & $  5.029\times 10^{-2}$ & $  6.135\times 10^{-2}$ & $  6.423\times 10^{-2}$ \\
Re187 & Re188m & $  2.181\times 10^{-2}$ & $  5.108\times 10^{-2}$ & $  5.264\times 10^{-2}$ \\
Pt192 & Pt191 & $  4.270\times 10^{-2}$ & $  4.238\times 10^{-2}$ & $  3.215\times 10^{-2}$ \\
Ir193m & Ir193 & $  2.090\times 10^{-3}$ & $  9.212\times 10^{-3}$ & $  2.514\times 10^{-2}$ \\
Ir191 & Ir192m & $  3.290\times 10^{-2}$ & $  1.785\times 10^{-2}$ & $  2.510\times 10^{-2}$ \\
W-186 & W-185m & $  4.033\times 10^{-2}$ & $  4.676\times 10^{-2}$ & $  2.323\times 10^{-2}$ \\
Re187 & Re186 & $  3.941\times 10^{-2}$ & $  2.248\times 10^{-2}$ & $  1.720\times 10^{-2}$ \\
Bi209 & Bi210 & $  6.132\times 10^{-5}$ & $  1.141\times 10^{-2}$ & $  1.679\times 10^{-2}$ \\
Ir192 & Ir193m & $  1.539\times 10^{-2}$ & $  1.425\times 10^{-2}$ & $  1.637\times 10^{-2}$ \\
Ir192 & Ir193 & $  1.845\times 10^{-2}$ & $  1.529\times 10^{-2}$ & $  1.352\times 10^{-2}$ \\
W-182 & W-181 & $  1.497\times 10^{-2}$ & $  1.685\times 10^{-2}$ & $  1.274\times 10^{-2}$ \\
Ir191 & Ir190 & $  2.775\times 10^{-2}$ & $  8.743\times 10^{-3}$ & $  1.263\times 10^{-2}$ \\
Pb208 & Pb207m & $  3.345\times 10^{-2}$ & $  9.360\times 10^{-3}$ & $  1.217\times 10^{-2}$ \\
Ir191 & Ir191m & $  2.715\times 10^{-2}$ & $  1.390\times 10^{-2}$ & $  1.167\times 10^{-2}$ \\
Pt194 & Pt193m & $  1.482\times 10^{-2}$ & $  1.318\times 10^{-2}$ & $  1.020\times 10^{-2}$ \\
W-186 & W-185 & $  3.379\times 10^{-2}$ & $  1.427\times 10^{-2}$ & $  9.663\times 10^{-3}$ \\
W-183 & W-183m & $  2.160\times 10^{-3}$ & $  1.927\times 10^{-2}$ & $  9.005\times 10^{-3}$ \\
W-183 & W-184 & $  5.385\times 10^{-3}$ & $  1.532\times 10^{-3}$ & $  8.812\times 10^{-3}$ \\
Ir193 & Ir192 & $  2.121\times 10^{-2}$ & $  8.499\times 10^{-3}$ & $  8.497\times 10^{-3}$ \\
Re185 & Re184m & $  1.874\times 10^{-2}$ & $  7.629\times 10^{-3}$ & $  8.467\times 10^{-3}$ \\
Ir194 & Ir195m & $  4.688\times 10^{-3}$ & $  5.884\times 10^{-3}$ & $  8.447\times 10^{-3}$ \\
Ir193 & Os193 & $  3.587\times 10^{-4}$ & $  2.355\times 10^{-3}$ & $  8.253\times 10^{-3}$ \\
Re188m & Re188 & $  1.010\times 10^{-2}$ & $  4.477\times 10^{-3}$ & $  7.839\times 10^{-3}$ \\
Bi210m & Bi210 & $  1.836\times 10^{-2}$ & $  3.068\times 10^{-3}$ & $  7.527\times 10^{-3}$ \\
\hline
\end{tabular}
\end{table}

\begin{table}
\caption{WMix case.
 Rankings for different methods.
}
\label{tab:dWMix}
\begin{tabular}{|p{2.0cm}|p{2.0cm}|c|c|c|}
\hline
& & \multicolumn{3}{|c|}{Sensitivity} \\
\hline
 Parent & Child & FDS & PBM & PRS \\
\hline
 Bi209 & Bi210 & $14$ & $38$ & $14$ \\
 Re187 & W-185m & $29$ & $0$ & $0$ \\
 Re187 & Re188m & $6$ & $12$ & $8$ \\
 Ir193 & Os191 & $35$ & $0$ & $0$ \\
 Ir193 & Os191m & $37$ & $0$ & $0$ \\
 Bi209 & Pb207m & $28$ & $0$ & $0$ \\
 Ir193 & Ir192m & $11$ & $0$ & $0$ \\
 Re187 & W-185 & $31$ & $0$ & $0$ \\
 W-184 & W-185m & $17$ & $50$ & $51$ \\
 Re187 & W-187 & $19$ & $0$ & $0$ \\
 Ir193 & Ir193m & $3$ & $16$ & $4$ \\
 Re187 & Ta183 & $32$ & $0$ & $0$ \\
 W-186 & W-185m & $7$ & $27$ & $12$ \\
 Re187 & Re186 & $8$ & $18$ & $13$ \\
 Re187 & Re188 & $1$ & $4$ & $1$ \\
 Pt194 & Os191 & $26$ & $0$ & $0$ \\
 Pt194 & Os191m & $30$ & $0$ & $0$ \\
 Pb208 & Pb207m & $12$ & $26$ & $19$ \\
 Re185 & W-185m & $20$ & $0$ & $0$ \\
 Ir193 & Ir192 & $9$ & $20$ & $26$ \\
 W-184 & W-185 & $4$ & $13$ & $6$ \\
 Ir193 & Ir194 & $2$ & $1$ & $2$ \\
 Pt194 & Ir193m & $34$ & $0$ & $0$ \\
 W-186 & W-185 & $10$ & $21$ & $22$ \\
 W-184 & Ta183 & $33$ & $0$ & $0$ \\
 W-184 & Ta182m & $38$ & $0$ & $0$ \\
 W-184 & Ta182 & $36$ & $0$ & $0$ \\
 W-184 & W-183m & $15$ & $35$ & $55$ \\
 Pt194 & Ir194 & $24$ & $0$ & $0$ \\
 Pt194 & Pt193m & $13$ & $24$ & $21$ \\
% Re185 & W-185 & $21$ & $0$ & $0$ \\
% W-186 & W-187 & $5$ & $8$ & $7$ \\
% Re185 & Re184m & $16$ & $46$ & $25$ \\
% Bi209 & Bi210m & $18$ & $61$ & $41$ \\
% Os188 & Re188m & $25$ & $0$ & $0$ \\
% Pt194 & Ir192 & $39$ & $0$ & $0$ \\
% Ir193 & Os193 & $23$ & $0$ & $28$ \\
% Os188 & W-185m & $27$ & $0$ & $0$ \\
% Ir191 & Os191 & $22$ & $0$ & $0$ \\
\hline
\end{tabular}
\end{table}

\begin{table}
\caption{WMix case.
 Rankings for different methods.
}
\label{tab:rWMix}
\begin{tabular}{|p{2.0cm}|p{2.0cm}|c|c|c|}
\hline
& & \multicolumn{3}{|c|}{Sensitivity} \\
\hline
 Parent & Child & FDS & PBM & PRS \\
\hline
 Ir193 & Ir194 & $2$ & $1$ & $2$ \\
 Re185 & Re186 & $0$ & $2$ & $5$ \\
 Ir191 & Ir192m & $0$ & $3$ & $11$ \\
 Re187 & Re188 & $1$ & $4$ & $1$ \\
 Ir191 & Ir192 & $0$ & $5$ & $3$ \\
 Ir192m & Ir192 & $0$ & $6$ & $71$ \\
 Ir192 & Ir193m & $0$ & $7$ & $15$ \\
 W-186 & W-187 & $5$ & $8$ & $7$ \\
 Ir193m & Ir193 & $0$ & $9$ & $10$ \\
 Ir192 & Ir193 & $0$ & $10$ & $16$ \\
 W-187 & Re187 & $0$ & $11$ & $0$ \\
 Re187 & Re188m & $6$ & $12$ & $8$ \\
 W-184 & W-185 & $4$ & $13$ & $6$ \\
 Re186 & W-186 & $0$ & $14$ & $37$ \\
 Re188m & Re188 & $0$ & $15$ & $29$ \\
 Ir193 & Ir193m & $3$ & $16$ & $4$ \\
 W-185 & Re185 & $0$ & $17$ & $0$ \\
 Re187 & Re186 & $8$ & $18$ & $13$ \\
 Ir194 & Ir195 & $0$ & $19$ & $62$ \\
 Ir193 & Ir192 & $9$ & $20$ & $26$ \\
 W-186 & W-185 & $10$ & $21$ & $22$ \\
 W-182 & W-181 & $0$ & $22$ & $17$ \\
 Pb207 & Pb207m & $0$ & $23$ & $54$ \\
 Pt194 & Pt193m & $13$ & $24$ & $21$ \\
 Ir194 & Pt194 & $0$ & $25$ & $0$ \\
 Pb208 & Pb207m & $12$ & $26$ & $19$ \\
 W-186 & W-185m & $7$ & $27$ & $12$ \\
 Ir191 & Ir191m & $0$ & $28$ & $20$ \\
 Re186 & Os186 & $0$ & $29$ & $0$ \\
 Os186 & Os185 & $0$ & $30$ & $38$ \\
\hline
\end{tabular}
\end{table}

\clearpage

\subsubsection{Y2O3}\label{sec:Y2O3}
This very simple case illustrates the ranking process in additional detail,
giving examples of the nuclear data that is used in the calculations.
Following irradiation, the activity at the end of this test case is dominated ($99.6$\,\%) by
two nuclides N-16 and Y-89m, see \Table{RNY2O3} for details.
The pathways analysis performed routinely by \textsc{Fispact-II} shows that all
pathways leading the $6$~nuclides listed, consist of just one reaction.
\begin{table}
\caption{Y2O3 case.
First five radio-nuclides in final inventory, ordered by activity,
plus Sr-89. Half-lives from the EASY-II database.
}
\label{tab:RNY2O3}
\begin{tabular}{|l|p{2.0cm}|c|c|c|c|}
\hline
Order & Nuclide & \multicolumn{2}{|c|}{Activity} & Atoms & Half-life \\
      &         & Bq  & Percent & & $\tau_k$  \\
\hline
1 & Y-89m & $2.5874 \times 10^{7}$ & $89.28\%$ & $5.847 \times 10^{8}$ & $15.663s$ \\
2 & N-16 & $2.9911 \times 10^{6}$ & $10.32\%$ & $3.077 \times 10^{7}$ & $7.13s$ \\
3 & Rb-86m & $1.0193 \times 10^{5}$ & $0.35\%$ & $8.971 \times 10^{6}$ & $1.017m$ \\
4 & C-15 & $1.0083 \times 10^{4}$ & $0.04\%$ & $3.562 \times 10^{4}$ & $2.449s$ \\
5 & Y-88 & $1.191 \times 10^{3}$ & $0.004\%$ & $1.583 \times 10^{10}$ & $107d$ \\
9 & Sr-89 & $62$ & $0.0002\%$ & $3.891 \times 10^{8}$ & $50.57d$ \\
\hline
\end{tabular}
\end{table}

\begin{table}
\caption{Y2O3 case.
Reactions identified as important by pathways analysis. Note that apart
from the reactions involving an excited parent species (labelled `m')
each reaction is equivalent to  a pathway. For the notation describing each
reaction see the \textsc{Fispact-II} manual.
}
\label{tab:reacY2O3}
\begin{tabular}{|p{3.5cm}|p{3.5cm}|p{3.5cm}|}
\hline
Pathway & Cross-section & $\epsilon_{ij}$ Relative   \\
Reaction & $barn$ & Uncertainty \%  \\
\hline
O-16(n,p)N-16  &  $0.03357$  &  $0.0$ \\
O-17(n,p)N-17  &  $0.01020$  &  $27.0$ \\
O-18(n,a)C-15  &  $0.05645$  &  $28.0$ \\
O-18(n,d)N-17  &  $8.9740\times10^{-6}$  &  $23.0$ \\
Rb-86m(b)Rb-86  &  $2.2675$  &  $4.39$ \\
Y-89(n,a)Rb-86  &  $0.0036110$  &  $70.0$ \\
Y-89(n,a)Rb-86m  &  $0.001820$  &  $70.0$ \\
Y-89(n,p)Sr-89  &  $0.02179$  &  $19.0$ \\
Y-89(n,2n)Y-88  &  $0.88630$  &  $5.7$ \\
Y-89(n,n)Y-89m & $0.4347$ & $8.5$ \\
Y-89(n,g)Y-90  &  $0.0022040$  &  $9.2$ \\
Y-89(n,g)Y-90m  &  $2.1410\times10^{-4}$  &  $9.2$ \\
Y-90m(n,g)Y-90  &  $2.4633$  &  $4.51$ \\
\hline
\end{tabular}
\end{table}

\begin{figure}[htbp]
\centerline{\rotatebox{270}{\includegraphics[width=9.0cm]{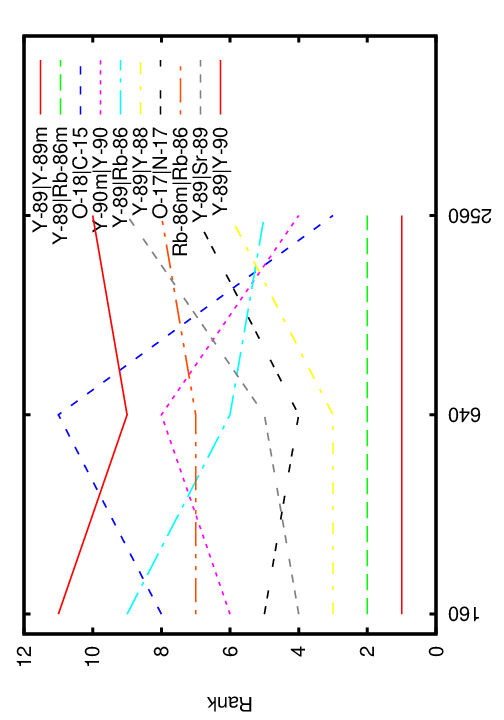}}}
\caption{\label{fig:cfY2O3_n3rank}
Comparison of the Y2O3 test case results, showing the first ten interactions
according to the scaled Pearson technique value~$S_{PRS}$, for a
Monte-Carlo sample size of~$N_x=2\,560$ per reaction, as $N_x$ is increased.}
\end{figure}

\begin{figure}[htbp]
\centerline{\rotatebox{270}{\includegraphics[width=9.0cm]{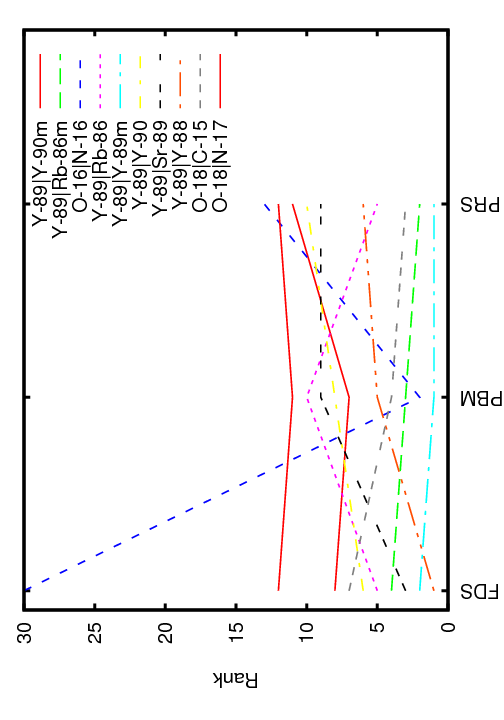}}}
\caption{\label{fig:cfY2O3_drank}
Comparison of the Y2O3 test case results, showing the first ten interactions
by magnitude of Fr\'{e}chet derivative. The labels are ordered according to
Fr\'{e}chet derivative size, so that the top interaction is the most sensitive.} 
\end{figure}

\paragraph{PRS} As a result of the dominance by two reactions, all the detailed rankings by Pearson
apart from the first two are suspect, see \Table{PY2O3} and
\Fig{cfY2O3_n3rank} in support of this contention. Moreover, no (zero) uncertainty
estimate is provided for the O-16$\vert$N-16 reaction, see \Table{reacY2O3}, the source of the N-16 in
the final inventory. Thus the Pearson
ranking is maximal, and in the other sections O-16$\vert$N-16 would have to be omitted from the
comparison plots.
Other important reactions are identified by inspection of comparison tables and plots such as
\Table{dY2O3} and \Fig{cfY2O3_drank} respectively. On this basis, the reaction
producing Sr-89 is identified as potentially important, hence its inclusion
in \Table{RNY2O3}.
\begin{table}
\caption{Y2O3 case.
{ Values of absolute Pearson correlation coefficient~$|r|$
as Monte Carlo sample size increases with~$N_x$.
}}
\label{tab:PY2O3}
\begin{tabular}{|p{2.0cm}|p{2.0cm}|c|c|c|}
\hline
& & \multicolumn{3}{|c|}{Absolute Pearson} \\ 
\hline
 Parent & Child & $160$ & $640$ & $2560$ \\ 
\hline
Y-89 & Y-89m & $  0.99945$ & $  0.99945$ & $  0.99948$ \\
Y-89 & Rb-86m & $  5.664\times 10^{-2}$ & $  4.483\times 10^{-2}$ & $  2.851\times 10^{-2}$ \\
O-18 & C-15 & $  1.768\times 10^{-3}$ & $  1.504\times 10^{-4}$ & $  8.698\times 10^{-3}$ \\
Y-90m & Y-90 & $  1.041\times 10^{-2}$ & $  1.250\times 10^{-2}$ & $  7.383\times 10^{-3}$ \\
Y-89 & Rb-86 & $  1.326\times 10^{-3}$ & $  1.010\times 10^{-2}$ & $  3.955\times 10^{-3}$ \\
O-18 & N-17 & $  2.644\times 10^{-2}$ & $  8.750\times 10^{-3}$ & $  3.389\times 10^{-3}$ \\
Y-89 & Y-90 & $  3.402\times 10^{-3}$ & $  6.061\times 10^{-3}$ & $  2.537\times 10^{-3}$ \\
Y-89 & Y-90m & $  1.537\times 10^{-2}$ & $  2.666\times 10^{-3}$ & $  2.261\times 10^{-3}$ \\
Rb-86m & Rb-86 & $  6.646\times 10^{-3}$ & $  1.651\times 10^{-2}$ & $  2.238\times 10^{-3}$ \\
O-17 & N-17 & $  1.821\times 10^{-2}$ & $  2.273\times 10^{-2}$ & $  1.766\times 10^{-3}$ \\
Y-89 & Y-88 & $  2.699\times 10^{-2}$ & $  1.402\times 10^{-2}$ & $  1.426\times 10^{-3}$ \\
Y-89 & Sr-89 & $  1.358\times 10^{-2}$ & $  1.085\times 10^{-2}$ & $  4.010\times 10^{-4}$ \\
O-16 & N-16 & $  2.585\times 10^{-16}$ & $  5.905\times 10^{-16}$ & $  1.552\times 10^{-15}$ \\
\hline
\end{tabular}
\end{table}

\clearpage
\paragraph{FDS} The Fr\'{e}chet derivative method works with a matrix that
has coefficients of all the possible reactions which involve the species
by the standard \textsc{Fispact-II} pathways analysis. There seems to be
little point in listing them all, obviously \Table{reacY2O3} is indicative.
As in the case of Pearson, the absence of an uncertainty estimate for the O-16$\vert$N-16 reaction
leads to a maximal ranking for the scaled Fr\'{e}chet derivative,
and in the other sections O-16$\vert$N-16 would have to be omitted from the
comparison plots.
\begin{table}
\caption{{ Y2O3 case.
 Rankings for different methods.
}}
\label{tab:dY2O3}
\begin{tabular}{|p{2.0cm}|p{2.0cm}|c|c|c|}
\hline
& & \multicolumn{3}{|c|}{Sensitivity} \\
\hline
 Parent & Child & FDS & PBM & PRS \\
\hline
 Y-89 & Y-90m & $8$ & $7$ & $11$ \\
 Y-89 & Rb-86m & $4$ & $3$ & $2$ \\
 O-16 & N-16 & $30$ & $2$ & $13$ \\
 Y-89 & Rb-86 & $5$ & $10$ & $5$ \\
 Y-89 & Y-89m & $2$ & $1$ & $1$ \\
 Y-89 & Y-90 & $6$ & $8$ & $10$ \\
 Y-89 & Sr-89 & $3$ & $9$ & $9$ \\
 Y-89 & Y-88 & $1$ & $5$ & $6$ \\
 O-18 & N-16 & $11$ & $0$ & $0$ \\
 O-18 & C-15 & $7$ & $4$ & $3$ \\
 O-18 & N-17 & $12$ & $11$ & $12$ \\
 O-17 & N-16 & $9$ & $0$ & $0$ \\
 O-17 & N-17 & $10$ & $6$ & $7$ \\
 Y-88 & Y-89m & $13$ & $0$ & $0$ \\
 O-16 & O-17 & $15$ & $0$ & $0$ \\
 Sr-89 & Y-89m & $31$ & $0$ & $0$ \\
 Rb-86 & Rb-86m & $14$ & $0$ & $0$ \\
 O-18 & O-16 & $21$ & $0$ & $0$ \\
 Y-90 & Y-90m & $18$ & $0$ & $0$ \\
 Y-90 & Rb-86m & $23$ & $0$ & $0$ \\
 O-18 & O-17 & $16$ & $0$ & $0$ \\
 O-17 & O-18 & $24$ & $0$ & $0$ \\
 Y-90 & Rb-86 & $22$ & $0$ & $0$ \\
 Y-90 & Y-89m & $19$ & $0$ & $0$ \\
 Y-90 & Sr-89 & $20$ & $0$ & $0$ \\
 O-17 & O-16 & $17$ & $0$ & $0$ \\
 Y-89m & Y-90m & $28$ & $0$ & $0$ \\
 Y-89m & Rb-86m & $26$ & $0$ & $0$ \\
 Y-89m & Rb-86 & $27$ & $0$ & $0$ \\
 Y-89m & Y-90 & $29$ & $0$ & $0$ \\
 Y-89m & Sr-89 & $25$ & $0$ & $0$ \\
\hline
\end{tabular}
\end{table}

\clearpage

\begin{figure}[htbp]
\centerline{\rotatebox{270}{\includegraphics[width=9.0cm]{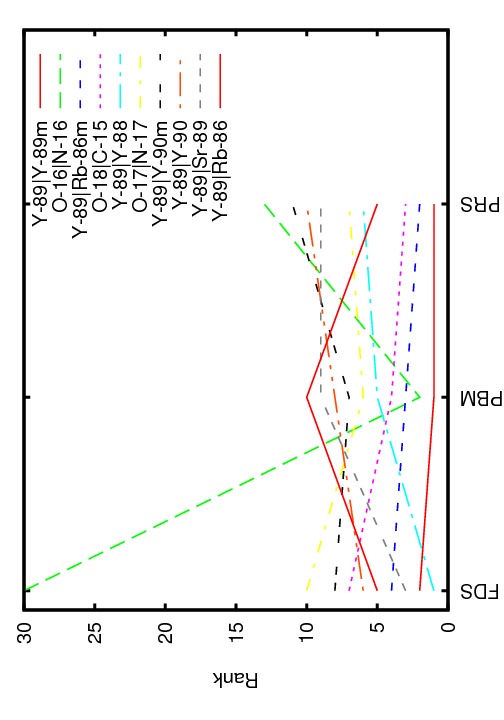}}}
\caption{\label{fig:cfY2O3_rrank}
Comparison of the Y2O3 test case results, showing the first ten interactions
according to the Pathways Based Metric~$S_{PBM}$.}
\end{figure}

\paragraph{PBM}
Since all pathways leading to the most important nuclides in terms
of activity, consist of just one reaction, it follows that these reactions
are simply ranked in order of contribution of the child species to the final activity.
Even though it has no associated uncertainty, the O-16$\vert$N-16 reaction can be ranked
by PBM. Taking into account the fact that  
in the other sections O-16$\vert$N-16 would have to be omitted from the
comparison plots, \Fig{cfY2O3_rrank} indicates that there is surprisingly
good agreement between the methods, even for reactions that contribute
little to the final activity.

\begin{table}
\caption{ Y2O3 case.
 Rankings for different methods.
}
\label{tab:rY2O3}
\begin{tabular}{|p{2.0cm}|p{2.0cm}|c|c|c|}
\hline
& & \multicolumn{3}{|c|}{Sensitivity} \\
\hline
 Parent & Child & FDS & PBM & PRS \\
\hline
 Y-89 & Y-89m & $2$ & $1$ & $1$ \\
 O-16 & N-16 & $30$ & $2$ & $13$ \\
 Y-89 & Rb-86m & $4$ & $3$ & $2$ \\
 O-18 & C-15 & $7$ & $4$ & $3$ \\
 Y-89 & Y-88 & $1$ & $5$ & $6$ \\
 O-17 & N-17 & $10$ & $6$ & $7$ \\
 Y-89 & Y-90m & $8$ & $7$ & $11$ \\
 Y-89 & Y-90 & $6$ & $8$ & $10$ \\
 Y-89 & Sr-89 & $3$ & $9$ & $9$ \\
 Y-89 & Rb-86 & $5$ & $10$ & $5$ \\
 O-18 & N-17 & $12$ & $11$ & $12$ \\
 Rb-86m & Rb-86 & $0$ & $12$ & $8$ \\
 Y-90m & Y-90 & $0$ & $13$ & $4$ \\
\hline
\end{tabular}
\end{table}

\clearpage
\section{Conclusions}\label{sec:conc}
The sensitivity of the total activity of an inventory to uncertainties in the nuclear data for
neutron-induced reactions has been studied. Six different test cases covering nearly the
whole range of atomic masses were considered using three different ranking techniques.
It is expected that similar results would be obtained for other inventory properties
and other projectile particle species.

The principal result is that a simple pathways-based metric~(PBM) gives a sensitivity ranking
of interactions which is comparable to ranking based on more conventional measures
obtained either by the direct method or in terms of Pearson correlation coefficients.
Moreover, the PBM is superior in that it
\begin{enumerate}
\item is quick to calculate once the principal pathways have been identified
\item does not suffer from numerical difficulties such as underflow (Fr\'{e}chet direct)
or round-off (Pearson) in its evaluation
\item may be generalised to the case of multiple irradiation periods just like
the pathways-reduced approach itself, whereas the other two techniques require further
investigation.
\item does not require error estimates for every interaction coefficient like Pearson.
\end{enumerate}

An additional noteworthy feature is that the PBM, which is a global measure
of uncertainty, is \emph{comparable} with more local measures, provided these
others are scaled by the uncertainty in the reaction cross-section. This scaling is
to be expected since the uncertainty estimates computed by \textsc{Fispact-II}~\cite[\S\,A.13]{ccfe-r11-11}
involve a multiplication by a measure of cross-section uncertainty (r.m.s.
is used to combine reaction coefficients rather than the simple percentages).
However, the product also involves the number of child nuclides in the inventory
which is a significantly different measure from the point sensitivity measures.

The value of studying a wide range of test cases is that it demonstrates the general
applicability of the above conclusions.
In conjunction with modifications to \textsc{Fispact-II}
for more efficient pathways-based analysis in the presence of multiple irradiations,
the PBM should be extended to account for loops in the pathways
and ultimately integrated into
a production version of the \textsc{Fispact-II} package.

\section*{Acknowledgement}\label{sec:ackn}
We are grateful to our colleagues J.-C.Sublet and J.W.Eastwood
for much advice and assistance.
%This work, part-funded by the European Communities under the contract of
%Association between EURATOM and CCFE, was carried out within the framework
%of the European Fusion Development Agreement. The views and opinions
%expressed herein do not necessarily reflect those of the European
%Commission. This work was also part-funded by the RCUK Energy Programme
%under grant EP/I501045.
%This work was funded by the RCUK Energy Programme under grant EP/I501045 and
%the European Communities under the contract of Association between EURATOM and CCFE.
%The views and opinions expressed herein do not necessarily reflect those of
%the European Commission.
%This work was funded by the RCUK Energy Programme grant number EP/I501045
%and the European Communities under the contract of Association between
%EURATOM and CCFE.
This work has been funded by the RCUK Energy Programme under grant number EP/I501045. 
To obtain further information on the data and models underlying this
paper please contact PublicationsManager@ukaea.uk.

The work of Relton and Higham was supported by European Research Council
Advanced Grant MATFUN~(267526)
and Engineering and Physical Sciences Research Council grant~EP/I03112X/1.

\section*{Appendix: Fr\'{e}chet Derivatives}\label{sec:appa}
As explained in \Sec{intro}, the Bateman equation~\Eq{rateq}
\begin{equation*}
  \frac{dX}{dt} = \mathsf{A}X,\quad X(0) = X_0,\quad t \in \left[0,t_f\right],
\end{equation*}
where $X \in \R^n$ is a vector of nuclide numbers and $\mathsf{A} \in \Rnbyn$
is a matrix of nuclear interaction coefficients,
controls the evolution of the nuclear activation over time.
In this appendix, we focus on the case where $\mathsf{A}$ is constant in time.

We are interested in the sensitivity of the total activity~\Eq{Q}
\begin{equation*}
  Q = \sum_{k = 1}^n \l_k X_k(t_f)
\end{equation*}
to the elements in $\mathsf{A}$,
which is determined by the $n^2$ numbers
$\partial Q / \partial A_{ij}$.
To determine these quantities we use the \mexp\ and its \Fd.
The \mexp\ of $\mathsf{A} \in \Rnbyn$ is defined by
\begin{equation*}
  e^\mathsf{A} = \sum_{k=1}^\infty \frac{\mathsf{A}^k}{k!}.
\end{equation*}
The \Fd\ of the exponential at $\mathsf{A}$ in the direction $\mathsf{E}\in\R^{\nbyn}$ 
is denoted by $\Lexp(\mathsf{A},\mathsf{E}) \in \Rnbyn$ and satisfies
\begin{equation*}
e^{\mathsf{A}+\mathsf{E}} = e^\mathsf{A} +
                             \Lexp(\mathsf{A},\mathsf{E}) +
                            o(\norm{\mathsf{E}}) \, .
\end{equation*}
For further details of \Fd s see \cite[Chap.~3]{high:FM}.

The solution to the Bateman equation is given by
\begin{equation*}
X(t) = e^{\mathsf{A}t}X_0
\end{equation*}
and so
\begin{equation*}
Q = f^T X(t_f) = f^T e^{\mathsf{A}t_f}X_0,\qquad
f = \left[\l_1 \dots\ \l_n\right]^T.
\end{equation*}
Let $\mathsf{E}_{ij}$ be the $\nbyn$ matrix with a $1$ in the $(i,j)$ entry and
zeros elsewhere.
Now,
\begin{align*}
  \frac{\partial{Q}}{\partial A_{ij}}
  &=
  \lim_{\delta \rightarrow 0}
  \frac{Q(A_{ij} + \delta) - Q(A_{ij})}{\delta}\\
  &=
  \lim_{\delta \rightarrow 0}
  \frac{f^T\left(e^{(\mathsf{A} + \mathsf{E}_{ij}\delta)t_f} -
                 e^{\mathsf{A}t_f}\right)X_0}{\delta}\\
  &=
  \lim_{\delta \rightarrow 0}
  \frac{f^T\left(L_{\exp}(\mathsf{A}t_f, \mathsf{E}_{ij}t_f\delta) +
                 o(\delta)\right)X_0}{\delta}\\
  &=
  t_f f^T L_{\exp}(\mathsf{A}t_f, \mathsf{E}_{ij})X_0,
\end{align*}
where we have used the fact that $\Lexp$ is linear in its second argument.

To determine the $k$ largest of these derivatives we can
simply compute them all and sort them.
For this we can use the relationship \cite[eq.~(3.16)]{high:FM}
\begin{equation}
  \label{eq.expm_relation}
  \exp\left(
    \begin{bmatrix}
      t\mathsf{A} & \mathsf{E}_{ij}\\
      0  & t\mathsf{A}
    \end{bmatrix}
  \right)
  =
  \begin{bmatrix}
    e^{t\mathsf{A}} & L_{\exp}(t\mathsf{A}, \mathsf{E}_{ij})\\
    0     & e^{t\mathsf{A}}
  \end{bmatrix},
\end{equation}
which yields the formula 
\begin{equation}
  \label{eq.expmv}
  \exp\left(
    \begin{bmatrix}
      t\mathsf{A} & \mathsf{E}_{ij}\\
      0  & t\mathsf{A}
    \end{bmatrix}
  \right)
  \begin{bmatrix}
    0\\
    X_0
  \end{bmatrix}
  =
  \begin{bmatrix}
    L_{\exp}(t\mathsf{A}, \mathsf{E}_{ij})X_0\\
    e^{t\mathsf{A}}X_0
  \end{bmatrix}.
\end{equation}
Hence one method to compute $L_{\exp}(\mathsf{A}t_f, \mathsf{E}_{ij})X_0$ is to apply
the method from \cite{alhi11} to compute the product on the \lhs\
and then read off the first $n$ components.

However, it is not necessary to carry out $n^2$ \Fd\ evaluations.
One suffices, as we now explain.
We need some notation.
{The Kronecker product of two matrices is now text-book for
numerical linear algebra, see e.g. ref~\cite{golubvanloan4}. It is defined
for matrices $\mathsf{B}$ and $\mathsf{C}$ (of any dimension)
as the block matrix $\mathsf{B} \otimes  \mathsf{C}  = (b_{ij}\mathsf{C})$.}
The $\vec$ operator stacks the columns of a matrix one of top of each
other from first to last, producing a long vector.
We need the property that 
$\vec(\Lexp(\mathsf{A},\mathsf{E})) = \mathsf{K}(\mathsf{A}) \vec(\mathsf{E})$, for some $n^2\times n^2$ matrix $\mathsf{K}(\mathsf{A})$
that satisfies $\mathsf{K}(\mathsf{A})^T = \mathsf{K}(\mathsf{A}^T)$.
Using the fact that the $\vec$ of a scalar is itself
and the formula
\begin{equation*}
\vec(\mathsf{A}\mathsf{X}\mathsf{B}) =
      (\mathsf{B}^T \otimes \mathsf{A})\vec(\mathsf{X}) \, ,
\end{equation*}
we have 
\begin{align*}
  \frac{\partial Q}{\partial A_{ij}}
  &=
  t_f f^T L_{\exp}(\mathsf{A} t_f, \mathsf{E}_{ij})X_0\\
  &=
  \vec\left(t_f f^T L_{\exp}(\mathsf{A} t_f, \mathsf{E}_{ij})X_0\right)\\
  &=
  t_f (X_0 \otimes f)^T \vec\left(L_{\exp}(\mathsf{A} t_f, \mathsf{E}_{ij})\right)\\
  & \equiv t_f g^T \mathsf{K}(\mathsf{A} t_f) \vec(\mathsf{E}_{ij}),
\end{align*}
where $g = X_0 \otimes f$.
Now, since $\vec(\mathsf{E}_{ij})$ is a unit vector, we simply require the $k$
largest elements in modulus of $g^T \mathsf{K}(\mathsf{A} t_f)$,
which are the largest $k$ elements in magnitude of $\mathsf{K}(\mathsf{A} t_f)^T g$.
We have $\mathsf{K}(\mathsf{A}^T t_f)g = \vec(L_{\exp}(\mathsf{A}^T t_f, \mathsf{E}))$,
where $\vec(\mathsf{E}) = g = X_0 \otimes f$ and hence $\mathsf{E} = fx_0^T$.
This means that a single \Fd\ evaluation is sufficient,
and it can be done using the relationship~\eqref{eq.expm_relation}
above with an \alg\ to compute the
\mexp\ such as that in~\cite{alhi09a}.

{
The computation of the exponential requires numerous matrix products, which
can occasionally cause numerical over- or under-flow due to the large range of
magnitudes in the coefficients arising in nuclear activation problems.
This may necessitate the use of quadruple precision arithmetic
on certain problems. In fact, quadruple precision was used to check the accuracy
of all the \Fd s calculated in the course of the current work.}

%\section*{References}

%\bibliographystyle{unsrt}
%\bibliography{waynes,misc,new,warv,neuts,active}
\end{document}